\documentclass[aps,prl,twocolumn,groupedaddress,epsfig]{revtex4}
\usepackage{graphicx,amssymb}

\begin{document}
\draft
\title{Non-equilibrium behavior of lysozyme solutions: beads, clusters and gels}
\author{H. Sedgwick$^{\ast}$, K. Kroy$^{\ast,\dagger}$, A. Salonen$^{\ast}$, M. B. Robertson$^{\ast}$, S. U. Egelhaaf$^{\ast}$ and W. C. K. Poon$^{\ast}$}

\date{\today}

\address{$^{\ast}$School of Physics, The University of Edinburgh,
Mayfield Road, Edinburgh EH9 3JZ, Scotland; $^{\dagger}$Hahn-Meitner-Institut Berlin, Glienicker Str. 100, 14109 Berlin, Germany}

\begin{abstract}

Observation of salt-induced aggregation of lysozyme at pH = 4.5, 22$^{\circ}$C  by optical microscopy revealed four regimes: bicontinuous texture, `beads', large aggregates, and transient gelation. The interaction of a metastable liquid-liquid binodal and an ergodic to non-ergodic transition boundary inside the equilibrium crystallization region can explain our findings. Lysozyme data at -2$^{\circ}$C, as well as many literature results, are consistent with the relative movement of these boundaries.

\end{abstract}
\pacs{PACS ?}

\maketitle

Short-range interparticle attractions are ubiquitous in synthetic colloids. Globular proteins in salt solution can also be modelled as `sticky particles' \cite{Rosenberger95,Tardieu99,Caccamo03}. The equilibrium phase behavior of particles with an isotropic short-range attraction is well known: the attractive interaction widens the fluid-crystal coexistence gap. When the interparticle potential is quantified by the second virial coefficient, a quasi-universal fluid-crystal coexistence boundary exists for synthetic colloids and proteins \cite{George94,Poon97,Poon00}. 

The non-equilibrium behavior of sticky particles is far less understood. In the case of colloids, it is known that two qualitatively different glassy states exist at volume fractions $\phi \gtrsim 0.6$ \cite{Pham02}. When $\phi \lesssim 0.2$ and the attraction is deep enough, permanent gelation is almost invariably observed. At shallower attractions ($\lesssim 10k_B T$), `transient' gelation and separation into particle-rich and particle-sparse amorphous phases have also been described \cite{ourgel,Poon99,Starrs02}, together with a recent report of a `cluster phase' \cite{Segre01}. Similar behavior occurs in protein solutions, although gels are less commonly reported. Theoretically, the gelation of sticky particles has long been discussed in terms of diffusion-limited cluster aggregation (DLCA) and percolation; recently mode-coupling theory (MCT) has been brought to bear on the problem \cite{Bergenholtz99}. It is also known that a metastable fluid-fluid phase boundary (binodal) exists. In polymer solutions, the interference of gelation and fluid-fluid phase separation leads to a rich `zoo' of non-equilibrium behavior \cite{Keller95,Jones01}. In this Letter, we show that this is also the case with particles.

\begin{figure}
\centering
\includegraphics[angle=0,width=9cm]{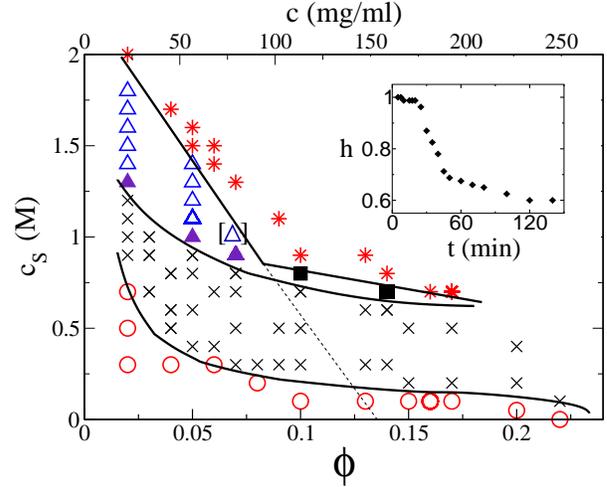}
\caption{Experimental phase diagram of lysozyme and NaCl at 22$^{\circ}$C and pH = 4.5. Axes give salt concentration, $c_s$, in molars, and lysozyme concentration in volume fraction, $\phi$ and weight/volume, $c$. Homogeneous solution ($\circ$); crystals ($\times$); bicontinuous texture ($\blacksquare$); spherical beads ($\blacktriangle$); large aggregates ($\vartriangle$); transient gels ($\ast$). Continuous lines estimate the positions of the crystallization boundary ({\it CB}, lowest), non-equilibrium boundary ({\it NEB}, middle) and gelation boundary (upper, piecewise straight), the low-$\phi$ portion of which extrapolates to $\phi \approx 0.13$ (thin line). Inset:  The height $h$ (scaled to 1 at $t=0$) of a transient gel ($\phi=0.07, c_s=1.4$~M) vs. time, $t$ (in minutes). The `latency time' is $\approx 30$~mins.} \label{phdiag}
\end{figure}

Our particles are the globular protein, lysozyme. Sodium chloride induces an effective short-range attraction by screening the intermolecular electrostatic repulsion \cite{Rosenberger95,Tardieu99,Caccamo03}. Careful visual observations and microscopy delineate four non-equilibrium regimes. We give a unified description of these observations within a framework that should be applicable to other sticky-particle systems. We show that the precise kind of phenomena observable in a particular system is sensitive to a range of physical parameters. This helps explain why a plethora of reported observations exists on apparently similar systems.

Six-times crystallized chicken egg-white lysozyme (Seikagaku America) was dissolved directly in a 50mM sodium acetate buffer titrated with hydrochloric acid to pH = 4.5. Stock solutions at 150~mg/ml were centrifuged to remove dust and undissolved material. Higher concentrations were reached using a Vivaspin 6 ml concentrator. Exact concentrations were determined by UV absorption spectroscopy after $1000\times$ dilution in buffer, using a specific absorption coefficient of 2.64~ml/mg.cm. (We calculate an effective $\phi$ using a molecular radius of 1.7~nm and molecular weight of 14320 g/mol.) Salt solution followed by deionized water were pipetted to the protein in buffer, mixed by shaking and observed by the naked eye. Small amounts were drawn into 0.1mm-thick rectangular capillaries and observed using a Zeiss optical microscope.

At increasing salt concentration, $c_s$, samples remained homogeneous ($\circ$, Fig.~\ref{phdiag}) until crystals ($\times$, Fig.~\ref{phdiag}) nucleated beyond a crystallization boundary (CB). At higher $c_s$, across a `non-equilibrium boundary' ({\it NEB}), all samples turned turbid after mixing. In the microscope, four regimes can be distinguished ($\blacksquare, \blacktriangle, \vartriangle$ and $\ast$, Fig.~\ref{phdiag}).

\begin{figure}
\centering
\includegraphics[angle=0,width=8cm]{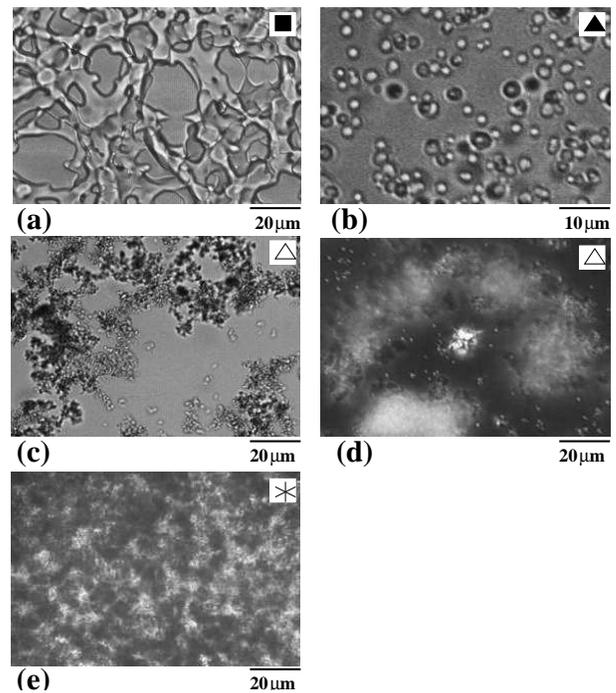}
\caption{Phase contrast images (symbols as in Fig.~\ref{phdiag}) showing (a) bicontinuous texture, ($\phi=0.14, c_s=0.7$~M); (b) spherical beads (note higher magnification), ($\phi=0.07, c_s=0.9$~M); (c) large aggregates, ($\phi=0.02, c_s=1.5$~M); (d) large aggregates and a crystal (bright, middle), ($\phi=0.05, c_s=1.1$~M), between crossed polarizers; (e) transient gel, ($\phi=0.1, c_s=0.9$~M), with inhomogeneities on the $\sim 10 \mu$m scale.} \label{micrographs}
\end{figure}

Macroscopically, samples with the highest $c_s$ ($\ast$, Fig.~\ref{phdiag}) turned white throughout after mixing. A plot of the height of the white portion of the sample vs. time showed a reverse sigmoidal shape (inset, Fig.~\ref{phdiag}). In other words, rapid sedimentation would start suddenly after a `latency period' ($\sim$ 15-30 minutes), Fig.~\ref{phdiag} inset. The resulting sediment crystallized after a few hours. Such `delayed sedimentation' is characteristic of `transient gels' \cite{Starrs02}. Note that our lysozyme transient gels are inhomogeneous on the $\sim 10 \mu$m ($\sim 10^3$-particle) level, Fig.~\ref{micrographs}(e).

Non-gelling samples above the {\it NEB} turned cloudy after mixing, and gave rise to turbid sediments without delay times. Although to the naked eye there is little to distinguish between the different samples in this region, microscopy revealed three distinct regimes. Just across the {\it NEB} and at $\phi \gtrsim 0.1$ ($\blacksquare$, Fig.~\ref{phdiag}), we observed bicontinuous textures reminiscent of spinodal decomposition, Fig.~\ref{micrographs}(a). Crystals subsequently nucleated exclusively in one of bicontinuous regions. Samples just across the {\it NEB} but with $\phi < 0.1$ ($\blacktriangle$, Fig.~\ref{phdiag}) showed freely-moving, non-coalescing spherical `beads' a few $\mu$m in diameter, or small aggregates of these beads, Fig.~\ref{micrographs}(b). Finally, at higher $c_s$ ($\vartriangle$, Fig.~\ref{phdiag}), we observed progressively larger aggregates; within each aggregate, structures on the scale of the previously-mentioned `beads' could be discerned, Fig.\ref{micrographs}(c). The beads and large aggregates would slowly dissolve once crystallization started within a few hours. 

The interaction between lysozyme molecules at high $c_s$ can be modelled by an isotropic, short-range attraction \cite{Rosenberger95,Tardieu99,Caccamo03}. Our observation of gelation at the highest $c_s$ ($\ast$, Fig.~\ref{phdiag}), which can be compared to the lysozyme gels observed by Muschol and Rosenberger \cite{Rosenberger97}, is therefore unsurprising. In an idealized system of particles with infinitely short-range and infinitely deep interparticle attraction, we expect irreversible DLCA to lead to a space-spanning gel \cite{Haw94,Hasmy96}. Gelation in real systems with finite-range and finite-depth attractions is well known, but its cause is intensely debated \cite{Gels}, a possible explanation being an ergodic to non-ergodic transition predicted by MCT \cite{Bergenholtz99} or a `renormalized' variant \cite{CMCT}. 

The non-permanent interparticle bonds give rise to ageing and, ultimately, collapse of the gel structure and rapid sedimentation after a `latency period'. Such `delayed sedimentation' is ubiquitous in synthetic colloids with short-range attraction (see \cite{Poon99,Starrs02} and references therein). After our  lysozyme gels collapsed, crystallization occurred. `Transient gelation' is therefore a metastable state {\it en route} to thermodynamic equilibrium.

The bicontinuous spinodal-type textures ($\blacksquare$, Fig.~\ref{phdiag} and \ref{micrographs}a) are also unsurprising. There is a metastable fluid-fluid binodal `buried' within the equilibrium fluid-crystal coexistence boundary of sticky particle systems \cite{Caccamo03,Miller03}, including lysozyme in salt solutions \cite{Rosenberger97,Zukoski03}. 

The `beads' and large aggregates are more intriguing. NMR measurements of these detected two populations of proteins: a fraction of free molecules (presumably in solution), and a fraction with severely restricted molecular motion. Preliminary X-ray diffraction gave no powder rings. Under crossed polarizers, the beads and large aggregates appeared birefringent, although less so than crystals, Fig.\ref{micrographs}(d). These results suggest that the beads and the large aggregates consist of lysozyme in an amorphous solid state. Our `beads' are reminiscent of the `oil-like droplets' frequently reported in studies of protein crystallization. A recent review \cite{McPherson01} noted that `crystals ... nucleate on the surfaces of [these] droplets \ldots and {\it they always grow outward} into the exterior \ldots protein containing medium, {\it not inward} into the \ldots protein rich phase.' This observation suggests that these `oil droplets' may in fact be amorphous solid phases on which heterogeneous nucleation takes place, consistent with our `beads'. 

\begin{figure}
\centering
\includegraphics[width=8cm]{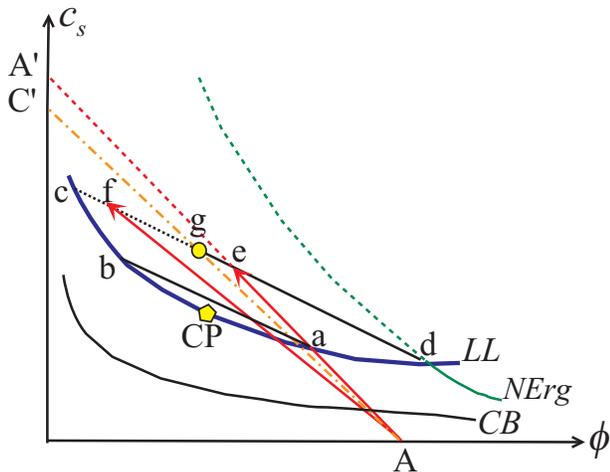}
\caption{Schematic illustration (in color on line) of the origins of observed nonequilibrium behavior for samples obtained by diluting protein solution with volume fraction $A$ with salt solution. {\it CB} = crystallization boundary, {\it NErg} = (underlying) ergodic to non-ergodic transition boundary, $LL$ = (metastable) liquid-liquid binodal with critical point (CP) and tie lines $ab$, $cd$. We predict gelation above $dg$-$gC^{\prime}$, large aggregates and beads in the region between $gC^{\prime}$ and $gc$, liquid-liquid phase separation below $dc$ but above $LL$, and crystallizaton between $LL$ and {\it CB}.} \label{scheme}
\end{figure}

We suggest that the formation of `beads' and large aggregates is due to the interaction of kinetic arrest and liquid-liquid phase separation. Our proposal is summarized in Fig.~\ref{scheme}. Here $CB$ is the equilibrium crystallization boundary and $LL$ is a liquid-liquid binodal. An ergodic to non-ergodic transition (`kinetic arrest') occurs across {\it NErg}, i.e., this is where gelation {\it would} occur in the absence of $LL$. The nature of this transition is debated  \cite{Bergenholtz99,Gels,CMCT}. At $\phi \gtrsim 0.4$, its position may be given rather accurately by MCT \cite{Pham02}. However, the precise identity of {\it NErg} does not affect what follows.

Now recall that a sample with (average) composition $e$ would have been obtained by mixing protein solution $A$ and brine $A^\prime$. We suggest that in this process the protein solution is `quenched' along the path $AA^{\prime}$ \cite{caveat}. When the quench path crosses $LL$ at $a$, phase separation into protein-rich and protein-poor solutions with compositions given by the ends of the tie line $ab$ occurs. If the quench is sufficiently slow compared to the rate of this phase separation, there will be time for {\it macroscopic} domains to develop and densify following the binodal along $ad$ and $bc$. In other words, a slow quench permits us to assume that a sample is always {\it in local thermodynamic equilibrium}. If the average composition is somewhere on $ae$, continuously coarsening bicontinuous texture results, until one of the phases crystallizes (cf. $\blacksquare$, Fig.~\ref{phdiag}).

If the average composition is $e$, the bicontinuous texture coarsens until the compositions of the phase-separated regions reach $c$ and $d$. Here, the continuous protein-rich region will become non-ergodic: $d$ is on {\it NErg}, and we find a gel with spatial inhomogeneities governed by the length scale of the bicontinuous texture at the point of gelation (cf. $\ast$, Fig.~\ref{phdiag} and Fig.~\ref{micrographs}(e)). 

Along the tie line $dc$ there exists a `dynamic percolation threshold' \cite{Binder87}, point $g$, to the left of which the phase-separated texture is no longer bicontinuous. Thus, for a quench path just to the left of $AgC^{\prime}$, we expect a texture that {\it just fails} to be connected through space. When the protein-rich portions of this {\it locally} bicontinuous texture becomes non-ergodic, large aggregates result ($\vartriangle$, Fig.~\ref{phdiag}). Paths further to the left of $AgC^{\prime}$ give more disconnected phase-separated domains and therefore smaller non-ergodic aggregates. Thus, quenching along a path from $A$ towards $f$ should give more or less isolated domains: protein-rich {\it droplets}. Their Ostwald ripening gives a peaked size distribution \cite{Lifshitz61}. When the composition of these droplets reaches $d$, they turn non-ergodic to give rather monodispersed `beads' ($\blacktriangle$, Fig.~\ref{phdiag}), whose mean size is determined by the interplay between the quench speed and the phase transition kinetics.

Thus, our prediction is that for all samples prepared by diluting protein solution $A$ with salt solution $A^{\prime}$, and at protein concentration below $d$, $dg$-$gC^{\prime}$ is the gelation boundary with $gC^{\prime}$ extrapolating back to $A$ on the protein axis, and $LL$ is the {\it NEB}. Experimentally, our gel boundary (separating $\ast$ from other symbols in Fig.~\ref{phdiag}) is consistent with this prediction: it is plausibly piecewise straight. The observed {\it NEB} (the upper boundary of $\times$, Fig.~\ref{phdiag}) indeed has the shape of a LL binodal. 

In detail, all our samples with $\phi \leq 0.1$ were prepared by diluting protein solution of volume fraction $\phi_0 = 0.132$ with brine. Consistent with Fig.~\ref{scheme}, the steeper part of the experimental gel boundary extrapolates to $\phi = 0.13 \approx \phi_0$.  Again, according to Fig.~\ref{scheme}, the shallower part of the gel boundary is part of a tie line of the LL binodal. If we assume this interpretation for the experimental gel boundary in Fig.~\ref{phdiag}, then two deductions follow. First, we can estimate (using the lever rule) that the sample at $\phi = 0.14, c_s = 0.7$~M (right-hand $\blacksquare$) should give $\sim 2/3$ high-protein-concentration domains, consistent with microscopic observations, Fig.~\ref{micrographs}(a). Secondly, we can estimate that the critical density $\phi_{\rm crit} \sim 0.12-0.13$, consistent with $\phi_{\rm crit} \sim 0.15$ under similar conditions \cite{Rosenberger97}.

Finally, quench-path dependence is a crucial part of our explanatory scheme: the observed gel boundary should change if the starting protein concentration (and therefore the quench path) is altered. A sample that gave large aggregates, $[\vartriangle]$, Fig.~\ref{phdiag}, gelled when prepared by diluting with brine a protein solution at $\phi = 0.176$. 

To summarize, we have observed different regimes of salt-induced aggregation and gelation in lysozyme solutions at pH = 4.5 and 22$^{\circ}$C. The existence of these regimes can be explained by a metastable liquid-liquid binodal and a non-ergodicity transition line intersecting within the equilibrium fluid-crystal coexistence region. 

\begin{figure}
\centering
\includegraphics[width=8cm]{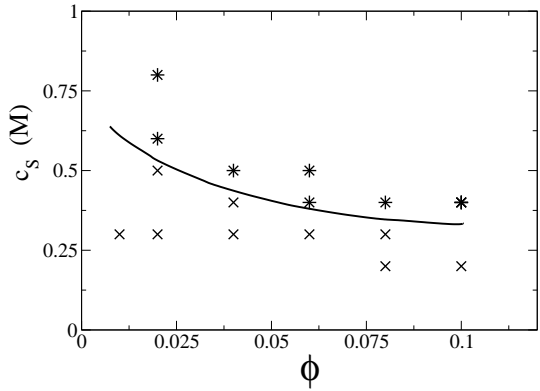}
\caption{Experimental phase diagram of lysozyme and NaCl at $-2^{\circ}$C and pH = 4.5: crystals ($\times$) and gels ($\ast$).} \label{temperature}
\end{figure}

It is probable that the relative positions of the various boundaries shown in Fig.~\ref{scheme}, and therefore the kinds of non-equilibrium behavior observable in an experiment, are rather sensitive to the precise parameters of a particular system (e.g. the range of the attraction \cite{shift}),  which in turn should be temperature dependent \cite{Tardieu99}. We repeated our experiments at temperatures -2$^{\circ}$C to 40$^{\circ}$C. Details will be reported elsewhere \cite{Sedgwick03}. Here we show results at -2$^{\circ}$C, Fig.~\ref{temperature}, where the {\it NErg} boundary has dropped below the {\it LL} binodal. Gelation was the only non-equilibrium behavior observed. 

We believe that the scenario described in Fig.~\ref{scheme} is also relevant to other experiments \cite{henksgel,ourgel,roberto} and simulations reported \cite{Nicolai02} in the literature . In many experiments, however, there are not enough published data points to justify this claim to the level of detail possible using the data in Fig.~\ref{phdiag}. More importantly, many data sets pertained to colloids with diameter in the $0.1 - 1 \mu$m range. The larger size slows down the rate of phase separation relative to the intrinsic particle dynamics (thus giving slower quenches). It also increases strongly the effect of gravity: clusters may sediment before they have time to grow large enough and touch to form a system-spanning gel. In some cases, charge may also be present, and the balance between a very short-range attraction and a longer range repulsion could generate non-trivial, new phenomena \cite{kegel}. A detailed study of charge and gravity effects will be presented elsewhere \cite{helen2}.

Thus, while there may be universality in the equilibrium phase behavior of particles with short-range attraction \cite{George94,Poon97,Poon00,Noro}, we have demonstrated that a bewildering `zoo' of non-equilibrium phenomena is possible, depending on precise conditions. A scheme involving kinetic pathways governed by the interaction between thermodynamic and kinetic transitions has been proposed that is able to capture and conceptualize this complexity.

We thank Lindsay Sawyer for X-ray diffraction and Mike Cates for discussions. KK was funded by the EU.


\begin{references}
\bibitem{caveat} Correspondingly, salt solution is `quenched' along the path $A^{\prime}A$; but the protein concentrations involved are lower, so that we do not discuss processes occurring in the salt-rich parts of our samples.
\end{references}
\end{document}